\documentclass[a4paper,10pt]{article}
\usepackage{ae}
\usepackage[ansinew]{inputenc}

\usepackage{lscape}
\usepackage{comment}
\usepackage{hyperref}
\usepackage{color}
\usepackage{amssymb,amsmath,amsthm,mathrsfs,graphicx}

\newcounter{mnotecount}[section]

\newcommand{\mnotex}[1]
{\protect{\stepcounter{mnotecount}}$^{\mbox{\footnotesize$\bullet$\themnotecount}}$ \marginpar{
\raggedright\tiny\em
$\!\!\!\!\!\!\,\bullet$\themnotecount: #1} }

\theoremstyle{plain}

\newtheorem{definition}{Definition}
\newtheorem{proposition}{Proposition}

\newtheorem{lem}{Lemma}

\def\eqq{\stackrel{\mathcal{H}}{=}}

 \title{Gravitational radiation and the evolution of gravitational collapse in cylindrical symmetry}
\author{Alfonso Garc\'ia-Parrado$^{\sharp}$\thanks{E-mail: alfonso@utf.mff.cuni.cz}~~and Filipe C. Mena$^{\flat\natural}$
\thanks{E-mail: fmena@math.uminho.pt}
\\\\
{\small $^\sharp$ Faculty of Mathematics and Physics, Charles University in Prague, V~Hole\v{s}ovi\v{c}k\'ach~2, 180~00 Praha 8, Czech Republic.}\\ 
{\small $^\flat$ Centro de Matem\'atica, Universidade do Minho, 4710-057 Braga,
Portugal}\\
{\small $^\natural$ Dep. Matem\'atica,}
{\small Instituto Superior T\'ecnico, Universidade de Lisboa, Av. Rovisco Pais 1, 1049-001 Lisboa, Portugal}}

\begin{document}

\maketitle

\begin{abstract}

Using the Sparling form and a geometric construction adapted to spacetimes with a 2-dimensional isometry group, 
we analyse a quasi-local measure of gravitational energy.
We then study the gra\-vi\-tational radiation through spacetime junctions in cylindrically symmetric models of gravitational 
collapse to singularities. The models result from the matching of collapsing dust fluids interiors with gravitational 
wave exteriors, given by the Einstein-Rosen type solutions.
For a given choice of a frame adapted to the symmetry of the matching hypersurface, 
we are able to compute the total gravitational energy radiated during the collapse and state whether 
the gravitational radiation 
is incoming or outgoing, in each case. This also enables us to distinguish whether a gravitational 
collapse is being enhanced by the gravitational radiation. 
\end{abstract}


\section{Introduction}
The theorem about the positivity of the global gravitational field at spatial infinity 
is a cornerstone of General Relativity (GR) and there are already different proofs available of this 
important result. In practical applications, 
it would be desirable to have a quasi-local notion of energy which could be applied to finite objects. A 
great deal of effort has been put towards this goal \cite{SZABADOS-REVIEW}, after the earlier 
proposals for the quasi-local mass by Hawking \cite{HAWKING} and Penrose \cite{PENROSE}. 
An interesting proposal was made by Brown and York \cite{BROWNYORK} using a Hamiltonian formulation of General Relativity, and
their definition depends on a gauge choice along three-dimensional spacelike hypersurfaces. Other authors have put forward 
definitions of mass which are based on more geometric constructions (see \cite{WANG-YAU,WANG2009} and references therein).

Associated to this problem is the problem of finding an appropriate Hamiltonian in each setting. 
In fact, the Hamiltonian for every diffeormorphism invariant theory depends on boundary conditions and is, 
therefore, non-unique. In General Relativity, for example, both the Komar Hamiltonian and 
the ADM Hamiltonian have been widely used (see e.g. \cite{CHRUSCIEL}).
The latter, in turn, is a particular case of the Von 
Freud super-potential \cite{FREUD} and a problem related to the search of 
superpotentials for GR is the question of the existence of the Lanzcos 
potential for the Weyl tensor (see e.g. \cite{EDGAR}).

An alternative approach to the definition of quasi-local gravitational energy quantities is given by the use 
of gravitational {\em pseudo-tensors}. In \cite{FRAUENDIENER-PSEUDO}, Frauendiener explains how to treat pseudo-tensors in the geometric
framework of the principal bundle defined by the frame bundle of a 4-dimensional Lorentzian manifold. Using this 
approach, it is shown that the energy-momentum pseudo-tensors of Einstein and Landau-Lifschitz can be recovered from pull-backs
to the spacetime manifold under appropriate sections of the so-called {\em Sparling 3-form}, which is defined 
in the bundle of linear frames. 
Szabados \cite{SZABADOS-PSEUDO}, generalizes these results and gives explicit formulae for gravitational pseudo-tensors  
in rigid basis or anholonomic frames. In particular, he shows that the pull-backs defined by coordinate sections  
of the contravariant and dual forms of the Sparling's form, defined on the bundle of linear frames over the spacetime, 
are the Bergman and the Landau-Lifshitz pseudo-tensors, respectively.

In most of the applications we are aware of, the quasi-local quantities are evaluated for an isolated system
at infinity, while in the present work we are interested in their local values at certain physically interesting 
hypersurfaces such as spacetime junctions. Even though quasi-local gravitational energy quantities are well-defined, 
they are frame dependent and, therefore, whenever they are used
it is necessary to give a justification about the choice of a particular frame. In the case of an isolated system, 
it is assumed that it is asymptotically flat and this means that one can introduce a group
of {\em asymptotic symmetries} which, roughly speaking, correspond to the symmetries of the 
Poincar\'e group defined in flat spacetime or to a generalization called the {\em BMS group}. 
These asymptotic symmetries induce {\em a set of privileged frames} in the spacetime, 
in the sense that they correspond asymptotically to the generators of the asymptotic symmetry group
and one can then take a frame of that set to compute a quasi-local gravitational energy quantity.
Within this approach, it is possible to recover the standard ADM and Bondi notions of momentum for an 
isolated system using definitions of quasi-local quantities based on pseudo-tensors \cite{GOLDBERG80}.
An approach based on the study of the asymptotic conditions as boundary conditions, defining when an
isolated system is radiating, is also possible \cite{TRAUTMAN}.

We would like to carry over the formalism described in the paragraph above 
to the case of a system composed by an (inhomogeneous) interior 
matched to a radiating exterior through a common hypersurface. The objective here is to give measures 
of the gravitational energy interchanged at the matching hypersurface, which plays now a similar role 
to the {\em spacelike} or {\em null infinity} of the case of an isolated system.
In the case of matching hypersurfaces with enough symmetry, there is a natural geometrical way to define frames
adapted to the symmetries of the hypersurface, and which will play the role of the frames corresponding to the 
asymptotic symmetries used to compute the quasi-local gravitational energy at infinity in an isolated system.

Here, to formulate this problem in a mathematically precise way, we use a frame bundle formalism and define equivalence 
classes between frames. We then apply our formalism to spacetimes admitting a $G_2$ isometry group and, in particular, 
calculate the radiated gravitational energy through the transitivity hypersurfaces of such isometry groups. 
A novel aspect of our work is the introduction of the notion of 
{\em gravitational collapse enhanced by the gravitational radiation}, i.e. 
by the gravitational energy-momentum flux, defined using the Sparling form. 
The meaning of this notion is that during the collapse, gravitational radiation is produced and 
the gravitational energy can either be radiated outside of the collapsing body or be radiated inwards, in which case it  
enhances the gravitational collapse.
Although our presentation is restricted to 4 dimensional spacetimes, most of it can be easily 
generalised to $n$ dimensions and can be applied e.g. to the brane world scenario where 4-dimensional 
branes are embedded, as hypersurfaces, on a 5-dimensional bulk.

Motivated by the recent detection of gravitational waves by LIGO \cite{LIGO}, an interesting application 
of our formalism would be to models of objects with a finite radius in astrophysics, emitting gravitational 
radiation into a vacuum exterior. Due to Birkhoff's theorem, it is not possible to have such vacuum exteriors 
in spherical symmetry. The next simplest symmetry assumption is cylindrical symmetry. So, we will consider models 
of gravitational collapse in cylindrical symmetry. Such models result from gluing an interior metric representing 
matter collapsing gravitationally to an exterior metric containing gravitational waves. An interesting aspect 
of these models is that the matter density is non-zero at the matching boundary, resulting in discontinuities in 
the Einstein field equations. Even though a realistic system undergoing gravitational collapse is not cylindrically symmetric, some non-trivial effects already
present in the real system can be found in the cylindrically symmetric model. For example, the degrees of freedom of the (pure) gravitational energy in the exterior 
may mix with the gravitational energy sourced by the matter at the boundary. These non-trivial effects may be 
captured by local, gauge-dependent, measures of gravitational energy.

The plan of the paper is as follows: in section \ref{sec:sparling}, we revise some properties of the Sparling 3-form and we compute 
its transformation under a change of frame.  In section \ref{sec:frame-equiv}, we review how 
to measure the gravitational energy radiated through embedded hypersurfaces of codimension 1 using the Sparling form. 
In section \ref{sec:gravitational-radiation}, we provide a geometric set-up 
for the definition of a frame based on the existence of
a submanifold representing the surface of a radiative body. 
In section \ref{cilinders}, we apply our formalism to cylindrically symmetric collapsing dust spacetimes 
containing gravitational radiation in the exterior which corresponds to the Einstein-Rosen solution.

Some algebraic computations of this paper have been done with {\em xAct} \cite{MARTINGARCIA}. 
We use Latin indices $a,b,c,..=1,2,3,4$, Greek indices $\alpha,\beta,..=1,2,3$, capitals $A,B,..=1,2$ and units such that $8\pi G=c=1$. 
\section{Sparling 3-form and Sparling identity}
\label{sec:sparling}

In this section and in the following section \ref{sec:frame-equiv}, we review some known notions and properties about  
the {\em Sparling form} and the 
{\em sparling identity} that are needed for our work. Even though most of the material is known, we 
present the equations in accordance to the notation that shall be followed in sections 
\ref{sec:gravitational-radiation} and \ref{cilinders}.

Let $(\mathcal{M},g_{ab})$ be a 4-dimensional spacetime with signature $(-,+,+,+)$ and consider 
the bundle of frames $L(\mathcal{M})$ over $\mathcal{M}$.
Let $B\equiv\{\vec{\boldsymbol e}_1,\vec{\boldsymbol e}_2,\vec{\boldsymbol e}_3,\vec{\boldsymbol e}_4\}$ be a 
chosen frame (smooth section of $L(\mathcal{M})$) and denote by 
$B^*\equiv\{{\boldsymbol\theta}^1,{\boldsymbol\theta}^2,{\boldsymbol\theta}^3,{\boldsymbol\theta}^4\}$ its dual co-basis. 
Define the {\em super-potential}
\begin{equation}
L(B,\vec{\boldsymbol e}_a)\equiv\frac{1}{2}\eta_{abcd}{\boldsymbol\theta}^d\wedge\Gamma^{bc}(B),
\end{equation}
where $\Gamma(B)^{b}_{\phantom{b}c}$ is the connection 1-form on the frame $B$ and $\eta_{abcd}$ 
are the components of the volume element of the 
metric $g_{ab}$ in the frame $B$. Note the dependency of the super-potential on both 
the frame $B$ and the frame elements 
$\vec{\boldsymbol e}_a$.
\footnote{One can define super-potential as a tensor valued form in the bundle of frames $L(\mathcal{M})$
as is done in \cite{FRAUENDIENER-PSEUDO,SZABADOS-PSEUDO}. To avoid introducing the notation and language 
used in fibre bundle theory, we adopt the point of view that the frame $B$ is a section of $L(\mathcal{M})$ and
we can use this section to pull-back the equations in $L(\mathcal{M})$ to $\mathcal{M}$. In this way, 
forms and tensors have their usual meaning of fields in the manifold $\mathcal{M}$, but the price payed 
for this simplification is the appearance of an explicit dependency on the frame $B$ in certain expressions 
(the {\em pseudo-tensors}).} 
Under the choice of the frame $B$, 
the super-potential
is a 2-form in $\Lambda^2(\mathcal{M})$ and it fulfills the {\em Sparling identity} 
\cite{FRAUENDIENER-PSEUDO,SZABADOS-PSEUDO,DUBOIS-VIOLETTE}
\begin{equation}
 d(L(B,\vec{\boldsymbol e}_a))=E(\vec{\boldsymbol e}_a)+S(B,\vec{\boldsymbol e}_a).
\end{equation}
where the quantity $E(\vec{\boldsymbol e}_a)\in\Lambda^3(\mathcal{M})$ is the so-called {\em Einstein 3-form}, which is given 
in terms of the curvature 2-form $R_{ab}$ by 
\begin{equation}
 E(\vec{\boldsymbol e}_a)=-\frac{1}{2}\eta_{abcd}{\boldsymbol\theta}^b\wedge R^{cd}\;,
\end{equation}
and it has the property
\begin{equation}
{\boldsymbol G}(\vec{\boldsymbol e}_a)\equiv *E(\vec{\boldsymbol e}_a)=G_{ab}{\boldsymbol\theta}^b
\in\Lambda^1(\mathcal{M})\;,
\end{equation}
with $*$ representing the Hodge dual and $G_{ab}$ the Einstein tensor. 

As $E(\vec{\boldsymbol e}_a)$ and  
$*E(\vec{\boldsymbol e}_a)$ are {\em tensorial} differential forms, we do not write their dependency on $B$,
such dependency being only kept for {\em pseudo-tensorial forms}. 
This is for instance the case of the 3-form $S(B,\vec{\boldsymbol e}_a)$ that it is given by
\begin{equation}
S(B,\vec{\boldsymbol e}_a)\equiv\frac{1}{2}(\eta_{bcde}{\boldsymbol\theta}^b\wedge\Gamma(B)^c_{\phantom{c}a}-
\eta_{abce}{\boldsymbol\theta}^b\wedge\Gamma(B)^{c}_{\phantom{c}d})\wedge\Gamma(B)^{de}.
\label{eq:sparling-form}
\end{equation}
We will follow the convention of \cite{FRAUENDIENER-PSEUDO,SZABADOS-PSEUDO} 
and call this form the {\em Sparling 3-form}.
For us, the important point about the Sparling identity is that it can be interpreted as a {\em conservation law} 
in the following way: By defining the 1-form
\begin{equation}
J(B,\vec{\boldsymbol e}_a)\equiv *dL(B,\vec{\boldsymbol e}_a),
\end{equation}
and computing the co-differential of $J(B,\vec{\boldsymbol e}_a)$, we get
\begin{equation}
\delta J(B,\vec{\boldsymbol e}_a)=-*d* J(B,\vec{\boldsymbol e}_a)=-*d**dL(B,\vec{\boldsymbol e}_a)=0.
\end{equation}
From this equation, we deduce that $J(B,\vec{\boldsymbol e}_a)$ always yields a conserved current, independently of the choice of the frame
$B$ or the frame vectors $\vec{\boldsymbol e}_a$. According to the above considerations, the current $J(B,\vec{\boldsymbol e}_a)$ can be split into 
two parts as
\begin{equation}
J(B,\vec{\boldsymbol e}_a)={\boldsymbol G}(\vec{\boldsymbol e}_a)+*S(B,\vec{\boldsymbol e}_a).
\end{equation}
The first part is a tensorial 1-form and, via the Einstein equations, can be interpreted as the flux of energy momentum due 
to the matter. 
The second part 
is non-tensorial and can be regarded as the flux of gravitational energy-momentum
with respect to the frame $B$.
If $\vec{\boldsymbol u}$ is a frame element of $B$ representing an observer, then the 1-form
\begin{equation}
\mathfrak{j}(B,\vec{\boldsymbol u})\equiv *S(B,\vec{\boldsymbol u})
\label{eq:gravitational-flux}
\end{equation}
shall be called the {\em gravitational energy-momentum flux} 1-form for the observer $\vec{\boldsymbol u}$ 
with respect to the frame $B$ or just the gra\-vi\-ta\-tional flux 1-form, if no confusion arises. 
It can be interpreted as the gravitational energy-momentum flux, which in this context is also called gravitational radiation, 
measured by the given observer in the 
frame $B$. In our context, we need to be able to decide if gravitational radiation is leaving or entering a radiating 
source. To that end, we put forward the following definition.
\begin{definition} We shall take the scalar quantity
\begin{equation}
 \mathfrak{j}(B,\vec{\boldsymbol u})\langle\vec{\boldsymbol e}_a\rangle 
\label{eq:radiation-flux}
\end{equation}
as the component of the gravitational radiation flux along the direction defined by the vector field
$(\vec{\boldsymbol e}_a)$. If this scalar quantity is positive (resp. negative), then we say that the gravitational radiation
is outgoing (resp. incoming) with respect to $(\vec{\boldsymbol e}_a)$.
\label{def:radiation-flux}
\end{definition}
The fact that $\mathfrak{j}(B,\vec{\boldsymbol u})$ is frame and observer dependent does not render it useless 
when computing the radiated gravitational energy in certain cases, as we illustrate through explicit examples in 
section \ref{cilinders}.
In fact, in \cite{FRAUENDIENER-PSEUDO}, it was proved that the scalar quantity (\ref{eq:radiation-flux})
has a correspondence with the components of the {\em energy-momentum complex} given by Einstein in 1916
\cite{EINSTEIN1916}.
Furthermore, it has been shown that, under an appropriate choice of frame $B$, it is possible to use this complex to 
recover the well-known notions of ADM and Bondi {\em momentum} at infinity for an isolated system 
(see \cite{GOLDBERG80}). Thus, following a similar 
approach, we will compute, in section \ref{cilinders},  
the gravitational energy radiated through a matching hypersurface separating two regions (exterior and interior) 
under the assumption that the hypersurface has certain symmetry 
properties. We will present explicit examples of systems undergoing gravitational 
collapse and show how the gravitational flux 1-form 
enables us to give a criterion stating whether gravitational radiation is 
being emitted or absorbed during the collapse.

It is obvious that a similar reasoning as the above can be performed if we take the Einstein form instead of the Sparling form. 
In this fashion, one can define the {\em matter current} with respect to the observer $\vec{\boldsymbol u}$ in the standard way
\begin{equation}
\mathfrak{J}(\vec{\boldsymbol u})\equiv {\boldsymbol G}(\vec{\boldsymbol u})\in\Lambda^1(\mathcal{M}).
\end{equation}
This current is tensorial and, as it is well-known, it represents the matter energy-momentum flux.
From previous considerations, it is clear that the sum of both currents is always conserved, a property that can be expressed 
by the condition 
\begin{equation}
\delta(\mathfrak{J}(\vec{\boldsymbol u})+\mathfrak{j}(B,\vec{\boldsymbol u}))=0.
\label{eq:energy-conservation}
\end{equation}
Interestingly, even though $\mathfrak{j}(B,\vec{\boldsymbol u})$ is frame dependent, the conservation property 
(\ref{eq:energy-conservation}) holds for any frame $B$. A particular case occurs when each of the currents are independently conserved.
As we will show, this is the situation for all the examples of section \ref{cilinders} modelling gravitational collapse. The implication of this is that, during the gravitational collapse, no fraction of the collapsing matter is transformed into gravitational radiation. 

\section{Frame equivalence and radiated gravitational energy through an embedded hypersurface} 
\label{sec:frame-equiv}
Suppose now that we introduce another frame $B'$ and its associate co-frame $B'^*$, related to $B$ and $B^*$ through
the formulae
\begin{equation}
\vec{\boldsymbol e}'_a=\gamma(B,B')_a^{\phantom{a}b}\vec{\boldsymbol e}_b\;,\quad 
{\boldsymbol\theta}'^a=\gamma(B',B)^{\phantom{b}a}_{b}{\boldsymbol\theta}^b\;,
\quad \gamma(B,B')_a^{\phantom{a}b}\gamma(B',B)^{\phantom{b}c}_{b}=\delta_a^{\phantom{a}c},
\label{eq:frame-transf}
\end{equation}
where the matrices $\gamma(B,B')$, $\gamma(B',B)$ encode the transition between the frames $B$ and $B'$. 
We regard these matrices as having entries which are scalars on $\mathcal{M}$ (0-forms). 
Under these circumstances, it is well-known that
the connection 1-form transforms as
\begin{equation}
\Gamma(B')^a_{\phantom{a}b}=\gamma(B',B)_q^{\phantom{q}a}\Gamma(B)^q_{\phantom{q}p}\gamma(B,B')_b^{\phantom{b}p}+
\gamma(B',B)_p^{\phantom{p}a}d(\gamma(B,B')_b^{\phantom{b}p}).
\label{eq:gamma-transform}
\end{equation}
We can use this expression to find the transformation law for the Sparling 3-form. A 
straightforward but tedious computation reveals:
\begin{lem} 
Under the change of frames $B$ to $B'$, the Sparling 3-form transforms as:
\begin{eqnarray}
\label{eq:sparling-transf}
&&S(B',\vec{\boldsymbol e}'_{s}) = - \tfrac{1}{2} \gamma(B,B')^{r \mathit{a}}
\eta_{\mathit{a} \mathit{b} \mathit{c} \mathit{d}} d \gamma(B,B')_{s}{}^{\mathit{b}} 
\wedge d \gamma(B,B')_{r}{}^{\mathit{c}} \wedge \theta^{\mathit{d}} -\nonumber\\
&&\tfrac{1}{2} \gamma(B',B)_{\mathit{c}}{}^{h}\gamma(B,B')_{s}{}^{\mathit{a}} 
\gamma(B,B')^{r \mathit{b}} \eta_{\mathit{a} \mathit{b} \mathit{d} \mathit{m}} d \gamma(B,B')_{r}{}^{\mathit{c}} \wedge d \gamma(B,B')_{h}{}^{\mathit{d}} \wedge \theta^{\mathit{m}} -\nonumber\\
&&\tfrac{1}{2} \eta^{\mathit{a}}{}_{\mathit{b} \mathit{c} \mathit{d}}
\Gamma(B)^{\mathit{b}}{}_{\mathit{a}} \wedge d \gamma(B,B')_{s}{}^{\mathit{c}} \wedge \theta^{\mathit{d}} -  
\tfrac{1}{2} \gamma(B',B)_{\mathit{b}}{}^{r} \gamma(B,B')_{s}{}^{\mathit{a}} \eta_{\mathit{a}}{}^{\mathit{c}}{}_{\mathit{d} \mathit{m}} \Gamma(B)^{\mathit{b}}{}_{\mathit{c}} 
\wedge d \gamma(B,B')_{r}{}^{\mathit{d}} \wedge \theta^{\mathit{m}} - \nonumber\\ 
&&\tfrac{1}{2} \gamma(B,B')_{s}{}^{\mathit{a}} \gamma(B,B')^{r \mathit{b}} \eta_{\mathit{b} \mathit{c} \mathit{d} \mathit{m}} \Gamma(B)^{\mathit{c}}{}_{\mathit{a}} \wedge 
d \gamma(B,B')_{r}{}^{\mathit{d}} \wedge \theta^{\mathit{m}} + \nonumber\\
&&\tfrac{1}{2} \gamma(B,B')_{s}{}^{\mathit{a}} \gamma(B,B')^{r \mathit{b}} 
\eta_{\mathit{a} \mathit{b} \mathit{c} \mathit{d}} \Gamma(B)^{\mathit{c}}{}_{\mathit{m}} \wedge d \gamma(B,B')_{r}{}^{\mathit{m}} \wedge \theta^{\mathit{d}} + 
\gamma(B,B')_{s}{}^{\mathit{a}} S(B,\vec{\boldsymbol e}_{\mathit{a}}). 
\end{eqnarray}
\end{lem}
From the expression above, it is clear that the relation between 
$S(B',\vec{\boldsymbol e}'_{s}) $ and $S(B,\vec{\boldsymbol e}_{\mathit{a}})$ 
is linear if $d \gamma(B,B')_{s}{}^{\mathit{a}}=0$
and, in that case, $S$ transforms like a tensor. We shall explore this fact next.

Let $\mathcal{H}\subset\mathcal{M}$ be a co-dimension 1 smooth embedded 
hypersurface and consider two frames $B$, $B'$ fulfilling the condition 
\begin{equation}
d(\phi^*(\gamma_{a}^{\phantom{a}b}(B',B)))=0\;,
\label{eq:phi-gamma}
\end{equation}
where $\phi:\mathcal{H}\rightarrow\mathcal{M}$ is a smooth embedding. 
Since one has that
\begin{equation}
 \phi^*(\gamma_{a}^{\phantom{a}b}(B',B))=\gamma_{a}^{\phantom{a}b}(B',B)|_{\mathcal{H}}
\label{eq:frame-transformation-on-H}
\end{equation}
we deduce that the scalars $\gamma_{a}^{\phantom{a}b}(B',B)|_{\mathcal{H}}$ can only be constants.
\begin{proposition}
For $\mathcal{H}\subset\mathcal{M}$ and $\phi:\mathcal{H}\rightarrow\mathcal{M}$ 
as described in the previous paragraph, the relation
\begin{equation}
 B\sim B' \Longleftrightarrow d(\phi^*(\gamma_{a}^{\phantom{a}b}(B',B)))=0
\end{equation}
is an equivalence relation in $\mathfrak{X}(L(\mathcal{M}))$ 
($\equiv$ the set of smooth sections in 
the frame bundle $L(\mathcal{M})$).
\end{proposition}
\proof
To ease the notation, we suppress indices in the matrix $\gamma_{a}^{\phantom{a}b}(B',B)$ and write it
simply as $\gamma(B',B)$. We show that the properties characterizing an equivalence relation are fulfilled:
\begin{itemize}
 \item Reflexive: As $\gamma(B,B)=I$ trivially $B\sim B$. 
 \item Symmetric: If $B\sim B'$ then $d(\phi^*\gamma(B',B))=0$. One has then
\begin{equation}
0=d(\phi^*(I))=d(\phi^*(\gamma(B',B)\phi^*(\gamma(B,B')))=\phi^*(\gamma(B',B))d(\phi^*(\gamma(B,B')).
\end{equation}
Since $\phi^*(\gamma(B',B))$ is invertible, the last equation entails $d(\phi^*(\gamma(B,B'))=0$ and thus
$B'\sim B$.
\item Transitive: if $B\sim B'$ and $B'\sim B''$ then 
\begin{eqnarray*}
&&d(\phi^*\gamma(B'',B))=d\big(\phi^*(\gamma(B'',B'))\phi^*(\gamma(B',B))\big)=0.
\end{eqnarray*}
Hence $B\sim B''$. \qed
\end{itemize}
\begin{proposition}
If there are two frames $B$, $B'$ on $\mathcal{M}$ such that
$B|_{\mathcal{H}}=B'|_{\mathcal{H}}$, then 
$B\sim B'$.
\label{prop:frame-sub-surface}
\end{proposition}
\proof If $B|_{\mathcal{H}}=B'|_{\mathcal{H}}$, then eq. \eqref{eq:frame-transformation-on-H} 
holds with $\gamma_{a}^{\phantom{a}b}(B',B)|_{\mathcal{H}}=\delta_a{}^{b}$ and therefore $B\sim B'$.\qed
\\\\
Now, let $B$, $B'$ be frames such that $B\sim B'$. Then
\begin{equation}
\phi^*(d(\gamma_{a}^{\phantom{a}b}(B',B)))=d(\phi^*(\gamma_{a}^{\phantom{a}b}(B',B)))=0. 
\end{equation}
Using this information in (\ref{eq:sparling-transf}), we deduce
\begin{equation}
\int_{\mathcal{H}} S(B',\vec{\boldsymbol e}'_a)=
\gamma(B',B)_a^{\phantom{a}b}|_{\mathcal{H}}\int_{\mathcal{H}}S(B,\vec{\boldsymbol e}_b)\;,
\label{eq:int-H-transformation}
\end{equation}
where we used the fact that $\gamma(B',B)_a^{\phantom{a}b}|_{\mathcal{H}}$ are constants
as $B\sim B'$. Bearing in mind this property, let us make the following definition
\begin{definition} 
Let $\mathcal{H}\subset\mathcal{M}$ be an embedded co-dimension 1 smooth hypersurface.
For a given frame $B$ and frame element $\vec{\boldsymbol e}_a\in B$, 
the gravitational flux through ${\cal H}$, with respect to $B$ and $\vec{\boldsymbol e}_a$ is defined as
\begin{equation}
 {\boldsymbol P}(B,\mathcal{H},\vec{\boldsymbol e}_a)\equiv\int_{\mathcal{H}} S(B,\vec{\boldsymbol e}_a).
\label{eq:define-P}
 \end{equation}
\end{definition}
If $B$ and $B'$ are frames such that $B\sim B'$, then eq. (\ref{eq:int-H-transformation}) can be rendered in the form
\begin{equation}
{\boldsymbol P}(B,\mathcal{H},\vec{\boldsymbol e}_a)=
\gamma(B',B)_a^{\phantom{a}b}|_{\mathcal{H}}{\boldsymbol P}(B',\mathcal{H},\vec{\boldsymbol e}'_b)\;,
\quad\text{for}\quad B\sim B'.
\end{equation}
This means that if we have a privileged frame $B$ in our problem, we can compute the gravitational
flux according to (\ref{eq:define-P}) and consider the scalars $\{{\boldsymbol P}(B,\mathcal{H},\vec{\boldsymbol e}_a)\}$ 
as the components of the gravitational 4-momentum of the hypersurface $\mathcal{H}$. 
We shall represent it as ${\boldsymbol P}(B,\mathcal{H})$.
 
An important particular case occurs in vacuum, when $S(B,\vec{\boldsymbol e}_a)=dL(B,\vec{\boldsymbol e}_a)$.
In this case, using Stokes theorem, we can write (\ref{eq:define-P}) as follows
\begin{equation}
{\boldsymbol P}(B,\mathcal{H},\vec{\boldsymbol e}_a)\equiv\int_{\mathcal{H}} dL(B,\vec{\boldsymbol e}_a)=
\int_{\partial\mathcal{H}} L(B,\vec{\boldsymbol e}_a).
\end{equation}
In an asymptotically flat isolated system we can choose $\mathcal{H}$ in such a 
way that one of the connected 
components of $\partial\mathcal{H}$ surrounds the source. Denoting by $\Sigma_r$ such 
a connected component,
we can study the existence of the limit
\begin{equation}
\lim_{r\rightarrow\infty}\int_{\Sigma_r} L(B,\vec{\boldsymbol e}_a)\;,
\label{eq:sigma-limit}
\end{equation}
for a hypersurface $\Sigma_r$ that approaches {\em infinity}. This analysis has been already
performed in a number of situations (see \cite{GOLDBERG80}):
\begin{enumerate}
 \item The surface $\Sigma_r$ approaches {\em spatial infinity}. In this case, if we choose 
an asymptotically Cartesian frame $B$, one can show that the limit (\ref{eq:sigma-limit})
corresponds to the ADM 4-momentum of the isolated source.
\item The surface $\Sigma_r$ approaches {\em null infinity}. In this case, if we choose a
frame $B$ as the {\em normalized translation} defined by the asymptotic BMS group, then 
the limit (\ref{eq:sigma-limit}) corresponds to the Bondi 4-momentum of the isolated source.
\end{enumerate}
In any case, we note that the choice of a frame $B$ with suitable properties at infinity, yields 
integral values of quasi-local quantities that have a valid physical meaning. In the next 
section, we shall follow this approach and show how the choice of a frame $B$
adapted to the symmetries of our physical system, at the matching hypersurface, 
enables us to analyse the problem of the gravitational radiation through the matching boundary.

The most important point to stress here is that the matching hypersurface, in our case, plays a role similar to 
{\em null infinity} or {\em spacelike infinity} in the case of an isolated system. As discused above, 
the choice of a frame $B$ adapted to the symmetries of null or spacelike infinity 
(the {\em asymptotic symmetries}) provides, respectively, the Bondi or ADM momentum. 
Therefore, a frame $B$ adapted to the symmetries of the matching hypersurface, should give us 
the quantities corresponding to the flux of gravitational energy-momentum through the matching hypersurface.
The detailed analysis of this assertion is presented in the next section.
 
\section{Gravitational flux through a hypersurface in a spacetime with a $G_2$ isometry group} 
\label{sec:gravitational-radiation}
Assume that $\mathcal{M}$ admits a maximal isometry group $G_2$ generated by the Killing 
vectors $\vec{\boldsymbol\xi}_1$, $\vec{\boldsymbol\xi}_2$. Let 
$\mathcal{H}\subset\mathcal{M}$ be a co-dimension 1 hypersurface. Choose a frame of the form
\begin{equation}
B=\{\vec{\boldsymbol n},\vec{\boldsymbol\xi}_1,\vec{\boldsymbol\xi}_2,\vec{\boldsymbol u}\}\;, 
\label{eq:type-frame}
\end{equation}
where $\vec{\boldsymbol n}$ is a vector field on $\mathcal{M}$ such that 
$\vec{\boldsymbol n}|_{\mathcal{H}}$ is unit and normal
to the hypersurface
and $\vec{\boldsymbol u}$ is an unit timelike vector field representing an observer. 
This frame is fully adapted to the geometry of the matching hypersurface $\mathcal{H}$ because its elements are either 
 fixed at $\mathcal{H}$ (as happens for an observer $\vec{\boldsymbol u}$ and the unit 
normal $\vec{\boldsymbol n}$) or are symmetries of the matching hypersurfaces (as happens for the 
Killing vectors $\vec{\boldsymbol\xi}_1$, $\vec{\boldsymbol\xi}_2$). 
There is a residual freedom in the choice of the frame
$B$, coming from the fact that the vector field $\vec{\boldsymbol n}$ 
is defined up to a sign on $\mathcal{H}$
and that
\begin{equation}
\vec{\boldsymbol\xi}'_1=\alpha^{11}\vec{\boldsymbol\xi}_1+\alpha^{12}\vec{\boldsymbol\xi}_2\;,\quad
\vec{\boldsymbol\xi}'_2=\alpha^{21}\vec{\boldsymbol\xi}_1+\alpha^{22}\vec{\boldsymbol\xi}_2\;,\quad
\alpha^{11}\;, \alpha^{12}\;, \alpha^{21}\;, \alpha^{22}\;\in\mathbb{R}\;,
\label{eq:killing-transf}
\end{equation}
are Killing vectors too. However, it is easy to get:
\begin{proposition}
Any other frame
$B'=\{\vec{\boldsymbol n},\vec{\boldsymbol\xi}'_1,\vec{\boldsymbol\xi}'_2,\vec{\boldsymbol u}\}$ with
$\vec{\boldsymbol\xi}'_1,\vec{\boldsymbol\xi}'_2$ related to $\vec{\boldsymbol\xi}_1,\vec{\boldsymbol\xi}_2$
by (\ref{eq:killing-transf}) fulfills the property 
$B'\sim B$.
\label{prop:G2frame}
\end{proposition}
From the previous proposition we deduce that, in a spacetime admitting a $G_2$ isometry group, one can define unambiguously
the gravitational 4-momentum of a hypersurface $\mathcal{H}$ as ${\boldsymbol P}(B,\mathcal{H})$, 
where $B$ is a frame of the form (\ref{eq:type-frame}). Since, in this case, the only freedom left to 
construct the frame $B$ is given by the choice of the observer $\vec{\boldsymbol u}$, we shall use the shorthand notation
\begin{equation}
{\boldsymbol P}(\vec{\boldsymbol u},\mathcal{H})\equiv {\boldsymbol P}(B,\mathcal{H}).
\label{eq:P-axial}
\end{equation}
Next, we are going to compute ${\boldsymbol P}(\vec{\boldsymbol u},\mathcal{H})$ in a number of practical examples.

\section{Gravitational flux in cylindrically symmetric spacetimes}
\label{cilinders}

In this section, we consider the problem of computing the gravitational flux
through hypersurfaces in cylindrically symmetric spacetimes. The hypersurfaces, here, 
result from the matching of an exterior vacuum spacetime containing gravitational waves 
and a dust fluid interior spacetime which is collapsing to form a singularity. To simplify the presentation, 
we consider exact solutions to the Einstein equations with diagonal metrics only.

We start by briefly recalling the matching conditions between two spacetimes $(\mathcal{M}^{\pm}, g^{\pm})$ across 
timelike hypersurfaces $\mathcal{H}^{\pm}$ (see more details in \cite{MARS-SENO}). The matching between 
two spacetimes requires an identification of their boundaries, i.e. a pair of embeddings
$\Phi^\pm:\; \mathcal{H} \longrightarrow \mathcal{M}^\pm$ with $\Phi^\pm(\mathcal{H}) =
\mathcal{H}^{\pm}$, where $\mathcal{H}$ is an abstract copy of one of the boundaries. Let $\xi^\alpha$ be 
a coordinate system on $\mathcal{H}$, where $\alpha,\beta=1,2,3$ are indices on the hypersurface. 
Given a vector basis $\{\partial/\partial \xi^\alpha\}$ of the tangent space $T\mathcal{H}$, 
the push-forwards $d\Phi^\pm$ provide a correspondence between the vectors of this basis and sets of 
linearly independent vectors tangent to
$\mathcal{H}^{\pm}$, given in appropriate coordinates by $e^{\pm a}_\alpha = \partial_{\xi^\alpha} \Phi^{\pm a}$.
Here, we follow the convention of previous sections that small Latin letters $a,b=1,2,3,4$ represent 
spacetime indices. There are also (up to orientation) vectors $n_{\pm}^{a}$ normal to the boundaries. 
The first and second fundamental forms on $\mathcal{H}$ are given by 
 $q^\pm=\Phi^{\pm\star}(g^\pm)$ and $K^\pm=\Phi^{\pm\star}(\nabla^\pm {\bf n^\pm})$, where 
$\Phi^{\pm\star}$ denotes the pull-back corresponding to the maps $\Phi^\pm$. In components, we may write
\begin{equation}
q_{\alpha\beta}^{\pm}\equiv e^{\pm a}_\alpha e^{\pm b}_\beta
g{}_{ab}|_{{}_{{\cal H}^\pm}},~~~
K_{\alpha\beta}^{\pm}=-n^{\pm}_{a} e^{\pm b}_\alpha\nabla^\pm_b e^{\pm
a}_\beta.
\end{equation}
The matching conditions between two spacetimes through $\mathcal{H}$ require the equality of the first and second fundamental form on $\mathcal{H}$, i.e.
\begin{equation}
  q_{\alpha\beta}^{+}=q_{\alpha\beta}^{-},~~~
  K_{\alpha\beta}^{+}=K_{\alpha\beta}^{-}.
\label{eq:backmc}
\end{equation}
In the examples below, the spacetimes are matched across cylinders of symmetry. 
\subsection{Gravitational wave exteriors}
The most general diagonal metric for cylindrically symmetric
vacuum spacetimes (with an Abelian $G_2$ acting on 2-dimensional spacelike hypersurfaces $S_2$) can be written as \cite{EXACT-SOL-BOOK}
\begin{equation} 
\label{ext}
g^{+}=e^{2(\gamma-\psi)}(-dT\otimes dT+d\rho\otimes d\rho)+R^2e^{-2\psi}d\tilde\varphi\otimes d\tilde\varphi+e^{2\psi} d\tilde z\otimes d\tilde z,
\end{equation}
where $\psi,\gamma, R$ are functions of the coordinates $\rho,T$ satisfying the Einstein equations \eqref{gammapsi}-\eqref{gammaT}. This metric is written in coordinates adapted to the Killing vectors 
$\partial_{\tilde\varphi}$ and $\partial_{\tilde z}$. From the physical point of view, the metric models cylindrical gravitational waves with one 
polarization state and, for the choice $R(T,\rho)=\rho$, corresponds to the metric found by Einstein and Rosen \cite{EINSTEIN-ROSEN}. So, we say that \eqref{ext} are metrics of {\em Einstein-Rosen type}. 

Dust matter sources to such spacetimes were studied in \cite{TOD-MENA, BRITO-ET-AL}, 
where the matching between collapsing interior spacetimes and Einstein-Rosen type of 
exteriors was shown to exist along timelike hypersurfaces. 
It would be therefore interesting to study the gravitational radiation through those hypersurfaces and to 
find out whether it enhances the gravitational collapse or not. 
It has been shown \cite{TOD-MENA} that, in those cases, the matching hypersurfaces are ruled by geodesics and, from the point of view of the exterior, 
can be parametrized as 
\begin{equation}
\mathcal{H}^+:\{t=\lambda, \rho=r_0, \tilde\varphi=\phi, \tilde z=\zeta\},
\end{equation}
where $r_0$ is constant and $\lambda$ is a parameter along the geodesics. 
So, $\mathcal{H}^+$ is generated by $\vec{\boldsymbol e}^+_1= \partial_t|_{\mathcal{H}^+}, 
\vec{\boldsymbol e}^+_2= \partial_{\tilde\varphi}|_{\mathcal{H}^+}$ and $\vec{\boldsymbol e}^+_3= \partial_{\tilde z}|_{\mathcal{H}^+}$, 
which together with $\vec{\boldsymbol n}^+=\partial_r|_{\mathcal{H}^+}$ 
form an orthogonal basis for the spacetime at $\mathcal{H}^+$.
\subsection{Spatially homogeneous interiors}
As interiors to \eqref{ext}, we first consider locally rotationally symmetric and spatially 
homogeneous dust solutions given by the Bianchi class $I$ of metrics \cite{EXACT-SOL-BOOK}, in 
cylindrical coordinates, as 
\begin{equation}
\label{compact}
g^{-}=-dt\otimes dt+a(t)^2 dz\otimes dz+b(t)^2(dr\otimes dr+r^2 d\varphi\otimes d\varphi)
\end{equation}
with
\begin{equation}
a(t)=(\alpha-t)^{-1/3}(\beta-t),~~~~
b(t)=(\alpha-t)^{2/3}
\end{equation}
for $0\le t<\rm{ min}\{\alpha,\beta\}$ and $r\le r_0$. 
If $\alpha =\beta$, the metric
reduces to the Friedman-Lema\^itre-Robertson-Walker (FLRW) class of spatially homogeneous and isotropic metrics. 

The interior metrics \eqref{compact} always
contain cylinders of symmetry which are trapped near the
singularity (see \cite{TOD-MENA}). The marginally-trapped cylinders trace out a
$3$-surface in the interior which eventually arrives at the
boundary of the matter located at $r=r_0$.

From the point of view of the interior, the matching hypersurface is parametrized as 
\begin{equation*}
\mathcal{H}^-:\{T=\lambda, r=r_0, \varphi=\phi, z=\zeta\},
\end{equation*}
so that $\mathcal{H}^-$ is generated by $\vec{\boldsymbol e}^-_1= \partial_T|_{\mathcal{H}^-}, 
\vec{\boldsymbol e}^-_2= \partial_{\varphi}|_{\mathcal{H}^-}$ and 
$\vec{\boldsymbol e}^-_3= \partial_{z}|_{\mathcal{H}^-}$, which together 
with $\vec{\boldsymbol n}^-=\partial_\rho|_{\mathcal{H}^-}$ form an orthogonal basis for the spacetime at $\mathcal{H}^-$. 

Then, the matching conditions \eqref{eq:backmc} between \eqref{ext} and \eqref{compact} across $\mathcal{H}$ give 
\begin{equation}
\label{match-conditions}
e^\psi  \eqq  a(t),~~~~ R \eqq  a(t)b(t) r,~~~~R_\rho  \eqq  a(t),~~~~\gamma\eqq\psi,~~~~\gamma_\rho\eqq\psi_\rho=0,
\end{equation} 
where $\eqq$ denotes evaluation at $\mathcal{H}$.
In order to solve the matching problem, the strategy followed in \cite{TOD-MENA} was as follows: For a suitable spacetime interior to \eqref{ext} and for the conditions \eqref{match-conditions} at the boundary, one obtains an explicit solution for $R(\rho,T)$ which satisfies the wave equation \eqref{EFE2} and
 the constraints \eqref{gammar} and
\eqref{gammaT}, at $\mathcal{H}$. In turn, the remaining
Einstein equations \eqref{gammapsi} and \eqref{wave} can be seen as providing
$\gamma_{,\rho\rho}$ and $\psi_{,\rho\rho}$ on $\mathcal{H}$.
Since we know data for the exterior metric and its normal
derivatives at the
boundary, it then follows \cite{TOD-MENA} that a unique $\psi$ exists on a neighbourhood ${\cal D}$ of $\mathcal{H}$. Since
$\gamma\eqq \psi$ and $\gamma_{,\rho}\eqq \psi_{,\rho}$, once we
have $\psi$, we use a similar argument in \eqref{gammapsi} to get
a unique $\gamma$ in ${\cal D}$.

\subsubsection{FLRW case}
\label{subsec:FLRW}
If the interior is given by a FLRW metric, then
$a(t)=b(t)=(\alpha-t)^{2/3}$, 
for $t\in (-\infty,\alpha)$. From the exterior, the matching surface $\mathcal{H}$ is the cylinder
$\rho=\rho_0,\, T<\alpha$, terminating in the singularity at
$T=\alpha$. At $\mathcal{H}$, we have
\begin{equation}
R  \eqq  r_0(\alpha-T)^{4/3},~~~~
R_{\rho}\eqq  (\alpha-T)^{\frac{2}{3}},
\end{equation}
so that, in particular, $R$ and $R_{\rho}$ on $\mathcal{H}$  are  positive for $T<\alpha$,
vanishing only at $T=\alpha$.
We may obtain $R$ explicitly as 
\begin{equation}
R=\frac{r_0}{2}(\alpha+\rho_0-T-\rho)^{\frac{4}{3}}-\frac{3}{10}(\alpha+\rho_0-T-\rho)^{\frac{5}{3}}+\frac{r_0}{2}(\alpha-\rho_0-T+\rho)^{\frac{4}{3}}+
\frac{3}{10}(\alpha-\rho_0-T+\rho)^{\frac{5}{3}} 
\end{equation}
and we may also get 
\begin{equation}
\psi  \eqq  \frac{2}{3}\ln{(\alpha-T)},~~~~~~\psi_{\rho}  \eqq  0. 
\end{equation}
We now wish to calculate \eqref{eq:P-axial} at $\mathcal{H}$ and, in order to do that, we need to choose an adequate frame
in the exterior.
Since $\psi=\gamma$ on $\mathcal{H}$, then
the holonomic frame $B=\{\partial_{\tilde\varphi}, \partial_{\tilde z}, \partial_\rho, \partial_T\}$ is such that
\begin{equation}
 \{\partial_T, \partial_\rho, \partial_{\tilde z}, \partial_{\tilde\varphi}\}|_{\mathcal{H}}
 =\left.\left\{e^{\gamma-\psi}\frac{\partial}{\partial T}, e^{\gamma-\psi}\frac{\partial}{\partial\rho},\vec{\boldsymbol\xi}_1,\vec{\boldsymbol\xi}_2\right\}\right|_{\mathcal{H}}\;,
\end{equation}
where the last frame has the properties of (\ref{eq:type-frame}). 
Proposition \ref{prop:frame-sub-surface} then implies 
that we can carry out our computations in the holonomic frame
with the choice $\vec{\boldsymbol u}=\exp({\gamma-\psi})\frac{\partial}{\partial T}$
as the vector field representing the observer.
In this case, a computation reveals that the Sparling 3-form at $\mathcal{H}$ is given by
\begin{equation}
\label{sparling-for-FLRW}
S^+_T|_{\mathcal{H}}=-\frac{2dT\wedge d\tilde z\wedge d\tilde\phi}{3(\alpha-T)^{\frac{1}{3}}},~~~~~
S^+_\rho|_{\mathcal{H}}=\frac{4 r_0 dT\wedge d\tilde z\wedge d\tilde\phi}{9(\alpha-T)^{\frac{2}{3}}},~~~~~S^+_{\tilde z}|_{\mathcal{H}}=0,~~~~~S^+_{\tilde \phi}|_{\mathcal{H}}=0.
\end{equation}
If we define now $\mathcal{H}^+\subset\mathcal{H}$ by the conditions $\tilde{z}_0\leq \tilde{z}\leq\tilde{z}_0+h$, $T_0\leq T\leq T_1$ 
with $\tilde{z}_0$, $T_0$, $T_1$ and $h$ constants, then one has
\begin{equation}
 \int_{\mathcal{H}^+} S^+_T=
2\pi h\left((\alpha-T_1)^{\frac{2}{3}}-(\alpha-T_0)^{\frac{2}{3}}\right)\;,\quad
\int_{\mathcal{H}^+} S^+_\rho=\frac{8\pi r_0 h}{3}\left((\alpha-T_0)^{\frac{1}{3}}-(\alpha-T_1)^{\frac{1}{3}}\right).
\end{equation}
From this, we can compute the four real numbers which define the 4-momentum ${\boldsymbol P}(\vec{\boldsymbol u},\mathcal{H}^+)$ 
(see eq. (\ref{eq:define-P})). Its components are
\begin{equation}
 {\boldsymbol P}(\vec{\boldsymbol u},\mathcal{H}^+)=\left(
 2\pi h\left((\alpha-T_1)^{\frac{2}{3}}-(\alpha-T_0)^{\frac{2}{3}}\right),
 \frac{8\pi r_0 h}{3}\left((\alpha-T_0)^{\frac{1}{3}}-(\alpha-T_1)^{\frac{1}{3}}\right),
 0,
 0
 \right).
\label{eq:pflrw-exterior}
\end{equation}
Also, we can compute the gravitational flux 1-form at the matching hypersurface
according to definition \ref{def:radiation-flux} and check whether the gravitational radiation is 
incoming or outgoing with respect to the radial direction $\partial/\partial\rho$. The result is
\begin{equation}
\mathfrak{j}(B,\vec{\boldsymbol u})\left\langle\frac{\partial}{\partial\rho}\right\rangle= 
\frac{2}{3 r_0(\alpha-T)^{\frac{5}{3}}}.
\label{eq:radiation-flux-FLR-exterior}
\end{equation}
It is interesting to note that the components 
of ${\boldsymbol P}(\vec{\boldsymbol u},\mathcal{H}^+)$ have a constant sign, 
namely, $P_T<0$ and $P_\rho>0$. Comparing with the sign of (\ref{eq:radiation-flux-FLR-exterior}), we deduce that
the gravitational energy-momentum flux 
is always outgoing. 
Even though one can also observe from \eqref{sparling-for-FLRW} that $S^+_T|_{\cal H}$, $S^+_\rho|_{\cal H}$ 
diverge as $T\to \alpha$, 
the flux integrals remain finite. This last conclusion is in line with 
the conclusions of \cite{TOD-MENA} using Weyl scalars.

As a consistency check, we can also carry out a similar computation for the interior, since the matching conditions 
entail $\mathcal{H}^+=\mathcal{H}^-=\mathcal{H}$ and 
${\boldsymbol P}(\vec{\boldsymbol u},\mathcal{H}^+)={\boldsymbol P}(\vec{\boldsymbol u},\mathcal{H}^-)$. 
To that end, we use the frame given by 
\begin{equation}
B=\left\{\partial_t, \frac{1}{b(t)}\partial_r, \partial_{z}, \partial_{\varphi}\right\}\;,
\end{equation}
because this is the frame which matches at $\mathcal{H}$ with the frame used for the exterior, namely
\begin{equation}
B=\{\partial_T, \partial_\rho, \partial_{\tilde z}, \partial_{\tilde\varphi}\}|_{\mathcal{H}}=
\left.\left\{\partial_t, \frac{1}{b(t)}\partial_r,  \partial_{z}, \partial_{\varphi}\right\}\right|_{\mathcal{H}} .
\end{equation}
Under this frame choice, the Sparling 3-form in the interior is
\begin{eqnarray}
S^-_t&=&-\frac{2}{3(\alpha-t)^{1/3}}dt\wedge dz\wedge d\phi+\frac{4r}{3(\alpha-t)^{2/3}}{\boldsymbol\theta}^1\wedge dz\wedge d\phi,\nonumber\\
S^-_r&=&\frac{4 r}{9(\alpha-t)^{2/3}}dt\wedge dz\wedge d\phi-\frac{4}{3(\alpha-t)^{1/3}}{\boldsymbol\theta}^1\wedge dz\wedge d\phi ,\nonumber\\
S^-_z&=&\frac{4 r}{3(\alpha-t)^{2/3}}dt\wedge {\boldsymbol\theta}^1\wedge d\phi\;,\quad
S^-_\phi= \frac{4 r}{3(\alpha-t)^{2/3}}dt\wedge {\boldsymbol\theta}^1\wedge dz\;,
\end{eqnarray}
where
\begin{equation}
 {\boldsymbol\theta}^1=(\alpha-t)^{2/3}dr.
\end{equation}
The gravitational flux 1-form in the interior is then given by
\begin{equation}
\mathfrak{j}\left(B,\frac{\partial}{\partial t}\right)=
{}^*{}_{g^-}(S^-_t)=\frac{1}{3 r (\alpha -  t)^2}(2 (\alpha -  t)dr-4 r dt ),  
\end{equation}
from which we get
\begin{equation}
\mathfrak{j}\left(B,\frac{\partial}{\partial t}\right)\left\langle\frac{1}{(\alpha-t)^{\frac{2}{3}}}\frac{\partial}{\partial r}\right\rangle
=\frac{2}{3 r(\alpha-t)^{\frac{5}{3}}}\;,
\label{eq:radiation-flux-FLRW}
\end{equation}
which agrees with (\ref{eq:radiation-flux-FLR-exterior}) at the matching hypersurface.
If we assume now that $\mathcal{H}^-\subset\mathcal{H}$ is defined by the conditions $z_0\leq z\leq z_0+h$, $T_0\leq t\leq T_1$ 
with $z_0$, $T_0$, $T_1$ and $h$ constants, then one has
\begin{eqnarray}
&&\int_{\mathcal{H}^-} S^-_t=2\pi h\left((\alpha-T_1)^{\frac{2}{3}}-(\alpha-T_0)^{\frac{2}{3}}\right)\;,\quad
\int_{\mathcal{H}^-} S^-_r=\frac{8\pi r_0 h}{3}\left((\alpha-T_0)^{\frac{1}{3}}-(\alpha-T_1)^{\frac{1}{3}}\right)\;,\quad\nonumber\\
&&\int_{\mathcal{H}^-} S^-_z= 0\;,\quad
\int_{\mathcal{H}^-} S^-_\phi= 0\;.\quad
\end{eqnarray}
Again, we compute the four real numbers which define the 4-momentum ${\boldsymbol P}(\vec{\boldsymbol u},\mathcal{H}^-)$ 
(see eq. (\ref{eq:define-P})) The result is
\begin{equation}
 {\boldsymbol P}(\vec{\boldsymbol u},\mathcal{H}^-)=\left(
 2\pi h\left((\alpha-T_1)^{\frac{2}{3}}-(\alpha-T_0)^{\frac{2}{3}}\right),
 \frac{8\pi r_0 h}{3}\left((\alpha-T_0)^{\frac{1}{3}}-(\alpha-T_1)^{\frac{1}{3}}\right),
 0,
 0
 \right).
\label{eq:pflrw-interior}
\end{equation}
Therefore, we can check explicitly that 
eqs. (\ref{eq:pflrw-exterior}) and (\ref{eq:pflrw-interior})
are consistent with ${\boldsymbol P}(\vec{\boldsymbol u},\mathcal{H}^-)
={\boldsymbol P}(\vec{\boldsymbol u},\mathcal{H}^+)$.

Since the interior is non-vacuum, we can also compute the Einstein 3-form with the result
\begin{equation}
E_t=-\frac{4}{3}r dr\wedge dz\wedge d\phi\;,\quad E_r=E_z=E_\phi=0.\;
\label{eq:einstein-current-flrw}
\end{equation}
From (\ref{eq:einstein-current-flrw}), we easily deduce that $dE_t=0$ which implies that $dS_t=0$, so the
matter energy-momentum current $\mathfrak{J}(\partial/\partial t)$ and $\mathfrak{j}\left(B,\partial/\partial t\right)$
are independently conserved currents.

\subsubsection{Bianchi $I$ case}
\label{subsec:BianchiI}
In this case, with equations 
\eqref{ext}-\eqref{match-conditions},
we may find \cite{TOD-MENA}
\begin{eqnarray}
R&=&\frac{r_0}{2}(\alpha+\rho_0-T-\rho)^{\frac{1}{3}}(\beta+\rho_0-T-\rho)-\frac{3}{4}(\alpha+\rho_0-T-\rho)^{\frac{2}{3}}(\beta+\rho_0-T-\rho)\nonumber\\
&&+
\frac{9}{20}(\alpha+\rho_0-T-\rho)^{\frac{5}{3}}
\nonumber\\
&&+\frac{r_0}{2}(\alpha-\rho_0-T+\rho)^{\frac{1}{3}}(\beta-\rho_0-T+\rho)+\frac{3}{4}(\alpha-\rho_0-T+\rho)^{\frac{2}{3}}(\beta-\rho_0-T+\rho)\nonumber\\&&-
\frac{9}{20}(\alpha-\rho_0-T+\rho)^{\frac{5}{3}}, \nonumber
\end{eqnarray}
as well as
 \begin{equation}
\psi\eqq\gamma\eqq \ln {(\beta-T)}-\frac{1}{3}\ln{(\alpha-T)}.
\end{equation}
Using this information, we get
\begin{eqnarray}
\left.S^+_T\right|_{\mathcal{H}}=\frac{(\beta-T)-3(\alpha-T) }{3 (\alpha -  T)^{4/3}}d T \wedge d \tilde{z} \wedge d \tilde{\phi}\;,\quad
\left.S^+_\rho\right|_{\mathcal{H}}=2 r_0\frac{ 3(\alpha-T)-(\beta-T)  }{9 (\alpha -  T)^{5/3}}d T \wedge d \tilde{z} \wedge d\tilde{\phi}\;,
\end{eqnarray}
the remaining components being zero. We define $\mathcal{H}^+\subset\mathcal{H}$ in a fashion similar as in subsection \ref{subsec:FLRW}
and perform the following computations
\begin{eqnarray}
&&\int_{\mathcal{H}^+} S^+_T=2\pi h\left[\frac{\beta-T_1}{(\alpha-T_1)^{\frac{1}{3}}}-\frac{\beta-T_0}{(\alpha-T_0)^{\frac{1}{3}}}
\right]\;,\\
&&\int_{\mathcal{H}^+} S^+_\rho=
\frac{2}{3}\pi h r_0\left[\frac{3 (\alpha-T_0) +(\beta -T_0)}{(\alpha -T_0)^{2/3}}
-\frac{3(\alpha-T_1) +(\beta -T_1)}{(\alpha -T_1)^{2/3}}
\right].
\end{eqnarray}
Hence, we deduce that in this case:
\begin{equation}
{\boldsymbol P}(\vec{\boldsymbol u},\mathcal{H}^+)=2\pi h\left(
\frac{\beta-T_1}{(\alpha-T_1)^{\frac{1}{3}}}-\frac{\beta-T_0}{(\alpha-T_0)^{\frac{1}{3}}},
\frac{r_0}{3}\left[\frac{3 (\alpha-T_0) +(\beta -T_0)}{(\alpha -T_0)^{2/3}}
-\frac{3(\alpha-T_1) +(\beta -T_1)}{(\alpha -T_1)^{2/3}}
\right],0,0
\right).
\label{eq:radiation-bianchi-I}
\end{equation}
Using the gravitational flux 1-form at the matching hypersurface, we get
\begin{equation}
\mathfrak{j}(B,\vec{\boldsymbol u})\left\langle\frac{\partial}{\partial\rho}\right\rangle={}^*{}_{g^+}(S^+_T)\left\langle\frac{\partial}{\partial\rho}\right\rangle
=\frac{(\beta-T)-3(\alpha-T)}{3 r_0(\alpha-T)^{\frac{5}{3}}(T-\beta)}. 
\label{eq:radiation-flux-BI-exterior}
\end{equation}
One can carry out a similar computation from the interior, as we did in subsection \ref{subsec:FLRW}, 
and use again the matching conditions to find that
${\boldsymbol P}(\vec{\boldsymbol u},\mathcal{H}^+)={\boldsymbol P}(\vec{\boldsymbol u},\mathcal{H}^-)$.
Also, the case $\alpha=\beta$ reduces to the FLRW case of the previous section.

Differently from the FLRW case, here one has to distinguish two cases: (i) $0<\alpha<\beta$, where the singularity is string-like along the axis of symmetry and $S_T$, $S_\rho$ 
diverge as $T\to \alpha$ and (ii) $0<\beta<\alpha$, where the spacetime singularity is pancake-like and $S_T$, $S_\rho$ are finite when $T\to \beta$. 
In the first case, the covector ${\boldsymbol P}(\vec{\boldsymbol u},\mathcal{H}^+)$ has components such that $P_T>0$ and $P_\rho<0$, whereas in the second case $P_T<0$ and $P_\rho<0$. 
Eq. (\ref{eq:radiation-flux-BI-exterior}) tells us that the gravitational flux with respect to the radial direction is incoming in the first case and outgoing in the second case.
In the first case, $P_T$ is ever increasing as the gravitational collapse evolves towards the singularity and, therefore, $P_T$ represents a net gain of gravitational energy whereas, in the second case $P_T$, 
measures a net loss of gravitational energy. 
From this, we deduce that, in the first case, the gravitational collapse is enhanced by incoming gravitational radiation, whereas in 
the second case gravitational radiation is outgoing as the gravitational collapse progresses.

For the Bianchi $I$ interior, a computation shows that (\ref{eq:einstein-current-flrw}) is also valid. 
Therefore, $dE_t=0$, which implies that $d S_t=0$, so again the
matter energy-momentum current $\mathfrak{J}(\partial/\partial t)$ and $\mathfrak{j}\left(B,\partial/\partial t\right)$
are independently conserved currents.
\subsection{Spatially inhomogeneous interior}
\label{subsec:SI}
The following metric corresponds to an inhomogeneous solution of the Einstein equations of the Szekeres family with a dust source and a regular axis \cite{SENO-VERA}:
\begin{equation}
\label{Senovilla-metric}
g^-=-dt\otimes dt+dr\otimes dr+\left(1-\frac{t^2+r^2}{\alpha^2}\right)^2
dz\otimes dz+r^2d\varphi\otimes d\varphi,
\end{equation}
where $\alpha \in \mathbb{R}\setminus\{0\}$. As the previous two cases, one can also show that this metric is an interior to Einstein-Rosen type metrics \cite{BRITO-ET-AL} and it is, thus, an interesting case to study by comparison with the spatially homogeneous cases.
  
In this case, the equality of the first fundamental forms on $\mathcal{H}$ gives
\begin{equation}
\label{gamma}
\gamma\eqq  \psi,~~~~
R  \eqq  r_0 \left(1-\frac{t^2+r_0^2}{\alpha^2}\right),~~~~
\psi  \eqq  \ln {\left(1-\frac{t^2+r_0^2}{\alpha^2}\right) },~~~~\gamma_{,T}\eqq \psi_{,T},
\end{equation}
while the equality of the second fundamental forms gives
\begin{equation}
\label{gammarho}
\gamma_{,\rho}\eqq  \psi_{,\rho},~~~~\psi_{,\rho} \eqq  -\frac{2r_0}{\alpha^2}\left(1-\frac{t^2+r_0^2}{\alpha^2}\right)^{-1},~~~~
R_{,\rho} \eqq 1-\frac{t^2+3r_0^2}{\alpha^2}.
\end{equation}
The solution to
\eqref{EFE2}, which satisfies the matching conditions at $\mathcal{H}$, is
\begin{eqnarray}
\label{Rsolution}
R(T,\rho)&=& r_0\left(1-\frac{r_0^2}{\alpha^2}\right) +\left(1-\frac{3r_0^2}{\alpha^2}\right)(\rho-\rho_0)-\frac{1}{6\alpha^2}\left((T+\rho-\rho_0)^3-(T-\rho+\rho_0)^3\right)\nonumber\\
&-&\frac{r_0}{2\alpha^2}\left((T-\rho+\rho_0)^2+(T+\rho-\rho_0)^2\right).\label{ext-R}
\end{eqnarray}
From the matching conditions $\psi\eqq\gamma$, we deduce that on $\mathcal{H}$
\begin{equation}
\{\partial_\rho, \partial_{\tilde\varphi}, \partial_{\tilde z}, \partial_T\}|_{\mathcal{H}}
 =\left.\left\{e^{\gamma-\psi}\frac{\partial}{\partial\rho},\vec{\boldsymbol\xi}_1,\vec{\boldsymbol\xi}_2,e^{\gamma-\psi}\frac{\partial}{\partial T}\right\}\right|_{\mathcal{H}}\;,
\label{eq:matching-frame-inhomogeneous}
\end{equation}
where the last frame has the properties of (\ref{eq:type-frame}). 
As we did in subsections \ref{subsec:FLRW} and \ref{subsec:BianchiI}, we appeal  to
Proposition \ref{prop:frame-sub-surface} to carry out our computations in the holonomic frame
$\{\partial_T, \partial_\rho, \partial_{\tilde\varphi}, \partial_{\tilde z}\}$, in order to compute \eqref{eq:sparling-form} on $\mathcal{H}$. In this case, we actually carry out the computation in all of the exterior with the result:
\begin{equation}
S^+_T=-\frac{2T}{\alpha^2}dT\wedge d\tilde{z}\wedge d\tilde{\varphi}\;,\quad
S^+_\rho=-\frac{2 r_0}{\alpha^2}dT\wedge d\tilde{z}\wedge d\tilde{\varphi}\;,\quad
S^+_{\tilde{z}}=S^+_{\tilde{\varphi}}=0.
\end{equation}
Next, we define $\mathcal{H}^+\subset\mathcal{H}$ and compute 
${\boldsymbol P}(\vec{\boldsymbol u},\mathcal{H}^+)$ as we did before. The result is
\begin{equation}
 {\boldsymbol P}(\vec{\boldsymbol u},\mathcal{H}^+)=\left(
\frac{4\pi h(T_0^2-T_1^2)}{\alpha^2}\;,
\frac{4 \pi h r_0}{\alpha^2}(T_0-T_1)\;,0\;,0
 \right).
\end{equation}
At the matching hypersurface, we have
\begin{equation}
\mathfrak{j}(B,\vec{\boldsymbol u})={}^*{}_{g^+}(S_T)=-\frac{2 T}{r_0(r_0^2-\alpha^2+T^2)}d\rho\;,
\quad
\mathfrak{j}(B,\vec{\boldsymbol u})\left\langle\frac{\partial}{\partial\rho}\right\rangle
=-\frac{2 T}{r_0(r_0^2-\alpha^2+T^2)},
\end{equation}
from which we deduce that the gravitational flux is outgoing through the matching hypersurface.
This indicates that the gravitational collapse is not enhanced by the gravitational radiation
at the matching hypersurface. 

In this particular case, we also supply an analysis of the collapsing process in 
the interior. To that end, we start by repeating the computation in the interior and find out whether the result agrees 
with what we have just found. In this case, the matching conditions entail
\begin{equation}
 \{\partial_T, \partial_\rho, \partial_{\tilde z}, \partial_{\tilde\varphi}\}|_{\mathcal{H}}=
\{\partial_t, -\partial_r, \partial_{z}, \partial_{\varphi}\}|_{\mathcal{H}}.
\end{equation}
Comparing with (\ref{eq:matching-frame-inhomogeneous}), we deduce that we can choose the frame
$\{\partial_t, -\partial_r, \partial_{z}, \partial_{\varphi}\}$ to carry out the 
computations in the interior. The result of these computations is
\begin{equation}
S^-_t=-\frac{2t}{\alpha^2}dt\wedge d{z}\wedge d{\varphi}\;,\quad
S^-_r=-\frac{2 r_0}{\alpha^2}dt\wedge d{z}\wedge d{\varphi}\;,\quad
S^-_{{z}}=S^-_{{\varphi}}=0.
\label{eq:sparling-form-inh-interior}
\end{equation}
So, this result yields
$$
{\boldsymbol P}(\vec{\boldsymbol u},\mathcal{H}^+)={\boldsymbol P}(\vec{\boldsymbol u},\mathcal{H}^-)\;,
$$
as expected.

The gravitational flux 1-form in the interior is explicitly given by
\begin{equation}
\mathfrak{j}^-(B,\vec{\boldsymbol u})={}^*{}_{g^-}(S_t)= \frac{2 ( rdt  + t dr)}{r (- \alpha^2 + r^2 + t^2)}
\;,
\end{equation}
To check whether the gravitational collapse is enhanced or not in the interior we choose to compute the gravitational radiation flux 
in the {\em apparent horizon} and, to that end, we analyse the trapped surface formation in the interior. 
In particular, we look at null 2-surfaces generated by the null
vectors $\vec k^{(\pm)}=\frac{\sqrt{2}}{2}(\partial_t\pm \partial_r)$. We then take the vectors $\vec e_1=\partial_\phi$ and $\vec e_2=\partial_z$, generators of the 2-cylinders and calculate the expansions
$\theta_{AB}^{(\pm)}= -k_a^{(\pm)} e_A^b \nabla_b e_B^a,$
where $A,B=1,2$, whose trace is denoted by $\theta^{(\pm)}$.
The condition $\theta^{(\pm)}=0$, for the existence of a {\em
marginally trapped cylinder}, is equivalent to
\begin{equation} 
\mathcal{T}:\;\;\label{ts1} t^2
\pm 2tr +3r^2=\alpha^2,
\end{equation}
for $t^2+r^2<\alpha^2.$ This means that,
for a given $\alpha$, there exists a positive $t^{(\pm)}_0=\mp
r+\sqrt{\alpha^2 -2r^2}$, for $r<\alpha/\sqrt{2},$ such that
$\theta^{(\pm)}=0$ and cylinders to the future of
$t^{(+)}_0$ are trapped. 

\begin{figure}[h]
\centering
    \includegraphics[width=11cm]{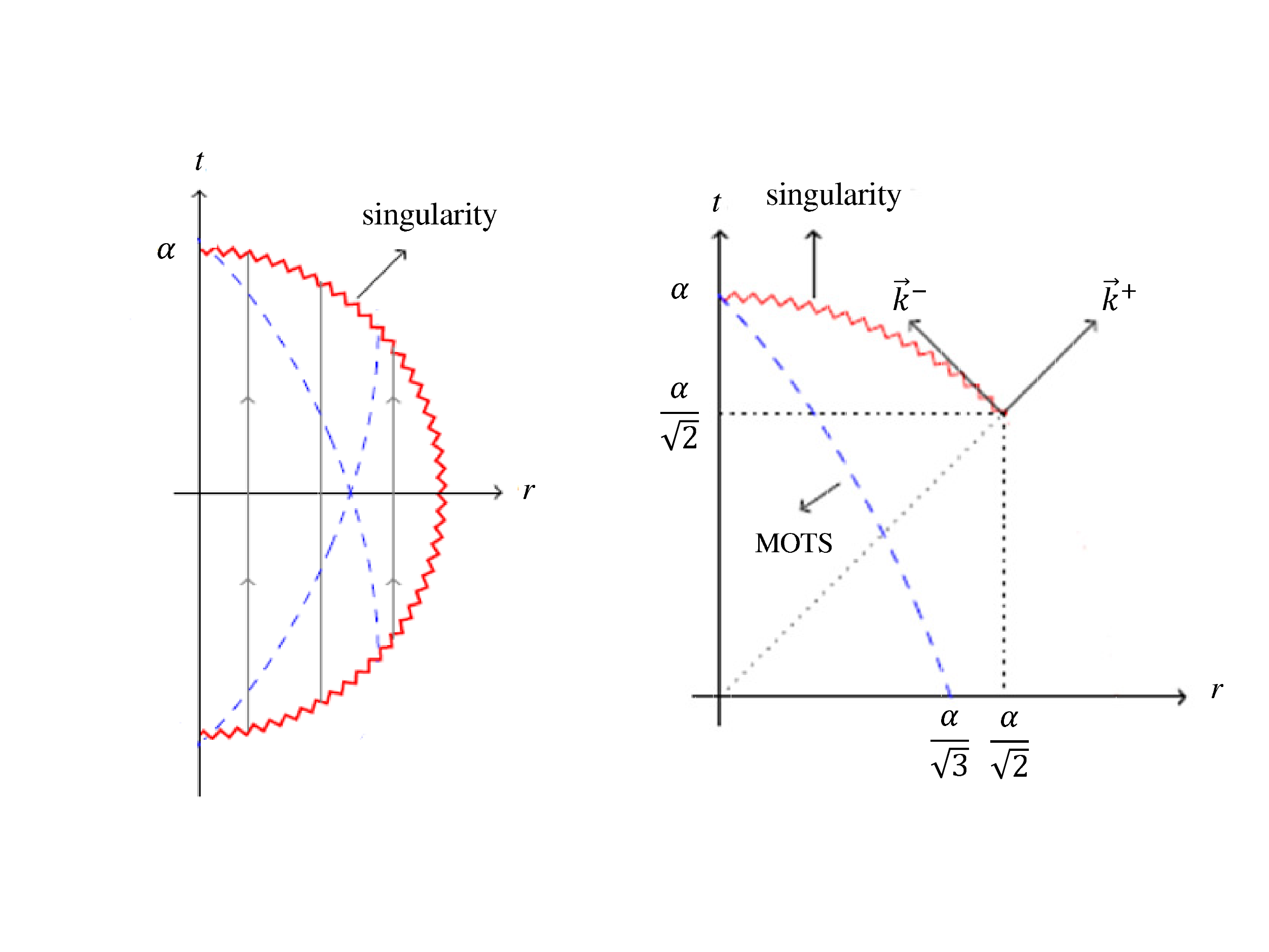}
  \caption{The left panel shows a diagram of the spacetime structure of (\ref{Senovilla-metric}), as obtained in \cite{SENO-VERA}.
  The vertical lines represent the geodesics along which the dust moves between singularities.
  The blue dashed lines represent the marginally trapped cylinders which are arcs of elipses. On the right panel, a cut of the spacetime is depicted
  for $r<\alpha/\sqrt{2}$. This figure is adapted from \cite{BRITO-ET-AL}.}\label{fig}
\end{figure}

To carry out a more detailed analysis, we pick the branch with a positive sign and  
introduce a new variable $x$ through the definition
$x\equiv t+ r$.
In terms of this new variable, condition (\ref{ts1}) adopts the form
$x^2+2r^2=\alpha^2$,
from which we conclude that in
the $(x,r)$ plane, condition (\ref{ts1}) represents arcs of
ellipses as shown in figure \ref{fig}.
We can now eliminate the variable $t$ from (\ref{eq:sparling-form-inh-interior}). Next,
using the map $x=\sqrt{\alpha^2-2r^2}$, we compute 
the pullback of the Sparling 3-form to the marginally trapped cylinders which yields
(here and in the following, we choose the positive sign in (\ref{ts1})):
\begin{eqnarray}
&&S^-_t= \frac{2 \left(\alpha ^2-4 r^2\right)}{\alpha ^2\sqrt{\alpha ^2-2 r^2}}dr\wedge d{z}\wedge d{\varphi}\;,\quad
S^-_r=\frac{2 \left(\alpha ^2- 2 r \left(\sqrt{\alpha^2-2 r^2}+ 2 r\right)\right)}{\alpha^2\sqrt{\alpha ^2-2 r^2}}dr\wedge d{z}\wedge d{\varphi}\;,\\
&& S^-_{{z}}=S^-_{{\varphi}}=0.
\label{eq:sparling-3-form-mots}
\end{eqnarray}
From these expressions, we can compute the gravitational flux 1-form on the marginally trapped cylinders using the 
standard definition, getting
\begin{equation}
\mathfrak{j}^-(B,\vec{\boldsymbol u})=\frac{dr}{r^2(2r^2-\alpha^2)}\left(\alpha^2-2r(2r+\sqrt{\alpha^2-2r^2})\right).
\label{eq:gravitational-flux-1-form-mots}
\end{equation}
We can also compute the gravitational energy-momentum flux through these cylinders
and we conclude that
\begin{equation}
 \int_{\mathcal{T}}S^-_t=\frac{4\pi h r_0 \sqrt{\alpha^2-2r_0^2}}{\alpha^2}\;,\quad
  \int_{\mathcal{T}}S^-_r=
  \frac{4\pi h r_0 \left(\sqrt{\alpha ^2-2 r_0^2}-r_0\right)}{\alpha^2}.
\label{eq:radiation-trapped}
\end{equation}
Since $r_0<\alpha/\sqrt{2}$ we deduce, from the first integral, that the observer will measure a net positive emitted gravitational energy.
However, looking at equation (\ref{eq:gravitational-flux-1-form-mots}), we deduce that the gravitational flux
with respect to $\partial/\partial r$
is incoming for $0<r<r_*$ and outgoing if $r_*<r<\alpha/\sqrt{3}$, where $r_*$ is given by
$$
r_*\equiv\frac{\alpha}{2} \sqrt{1-\frac{1}{\sqrt{3}}}.
$$
Therefore, when the singularity is approached the radiation is incoming but when one gets towards the matching hypersurface the radiation 
is outgoing, consistently with the analysis carried out in the exterior. This means that the gravitational collapse will 
be enhanced by the gravitational radiation when the singularity is approached ($r<r_*$) but not when $r>r_*$ (in particular, not 
in the matching hypersurface between the interior and the exterior).

To finish, we compute the Einstein 3-form in the interior obtaining
\begin{equation}
E_t=-\frac{4}{\alpha^2}r dr\wedge dz\wedge d\phi\;,\quad E_r=E_{{z}}=E_{{\phi}}=0.\;
\label{eq:einstein-current-inhom}
\end{equation}
From this result, we deduce that $dE_t=0$ which entails $dS_t=0$, so in this case the
matter energy-momentum current $\mathfrak{J}^-(\partial/\partial t)$ and $\mathfrak{j}^-\left(B,\partial/\partial t\right)$
are also independently conserved currents in the interior.

\section{Conclusions}
\label{sec:conclusions}

We have considered a quasi-local measure of gravitational energy, using the Sparling 3-form and a geometric construction adapted to spacetimes with a 2-dimensional isometry group. We have then studied the gravitational energy-momentum flux in models of gravitational collapse with cylindrical symmetry.
 
Taking advantage of the existence of the Killing vectors, we defined a frame adapted to our problem to compute the gravitational radiation at the boundary and at the interior of the collapsing body. The interiors we have analysed contain a dust fluid and are 
 FLRW, Bianchi $I$ and Szekeres solutions, whereas the exterior is always 
a vacuum Einstein-Rosen type solution containing gravitational waves. Our method shows that in the collapse modelled with the FLRW and inhomogeneous interiors, 
the gravitational radiation is always outgoing at the matching boundary of the collapsing body whereas, in the case of the Bianchi $I$, the gravitational radiation can be 
either outgoing
or incoming at the boundary during the collapse. 

We find that, in a model whose collapsing interior is a Bianchi $I$ spatially homogeneous 
spacetime with a string like singularity, the collapse is being enhanced by the gravitational radiation coming from the exterior.
 In the other cases analysed, the gravitational radiation is outgoing from the matching hypersurface during the collapse process. 
Note that these considerations pertain only to the analysis at the matching hypersurface. 

In the case of the inhomogeneous interior, we also carried out the analysis at the apparent horizon and found that, 
there, the gravitational collapse is enhanced 
towards its final phase, whereas it is not so at earlier stages. In this case, since the collapse 
is not enhanced at the matching hypersurface, we conclude that a gravitational energy-momentum flux 
is originated in the interior enhancing the collapse during its late evolution.

Our results show how a quasi-local measure of gravitational energy can be constructed in a geometrical
way using the Sparling form, and can be applied to the problem of gravitational collapse of a fluid body 
having an exterior with gravitational waves.

\section*{Acknowledgments}
We thank an anonymous referee for constructive criticisms on an earlier version of this paper.
We thank  FCT Projects Est-OE/MAT/UI0013/2014, PTDC/MAT-ANA/1275/2014 and CMAT, Univ. Minho, 
through FEDER Funds COMPETE. We also thank the Erwin Schr\"odinger International Institute 
for Mathematical Physics, ESI, where part of this work has been done. FCM thanks FCT for grant SFRH/BSAB/130242/2017.  
AGP thanks the financial support from Grant 14-37086G of the Czech Science Foundation and the
partial support from the projects IT956-16 (``Eusko Jaur\-la\-ri\-tza'', Spain),
FIS2014-57956-P (``Ministerio de Econom\'{\i}a y Competitividad'', Spain).
\section{Appendix}
The EFEs for the cylindrically symmetric vacuum metric (\ref{ext})
are
\begin{eqnarray}
\label{gammapsi}
0&=&\gamma_{,TT}-\gamma_{,\rho\rho}-\psi_{,\rho}^{2}+\psi_{,T}^{2}\\
\label{EFE2}
0&=&R_{,TT}-R_{,\rho\rho}\\
\label{wave}
0&=&\psi_{,TT}+\frac{R_{,T}}{R}\psi_{,T}-\psi_{,\rho\rho}-\frac{R_{,\rho}}{R}\psi_{,\rho}
\end{eqnarray} together with the two constraint equations \begin{eqnarray}
\label{gammar} \gamma_{,\rho} & = & \frac{1}{R_{,\rho}^{2}-
R_{,T}^{2}}(RR_{,\rho}(\psi_{,T}^{2}+\psi_{,\rho}^{2})-2RR_{,T}\psi_{,T}\psi_{,\rho}+R_{,\rho}
R_{,\rho\rho}-R_{,T}R_{,T\rho})\\
\label{gammaT} \gamma_{,T} & = & -\frac{1}{R_{,T}^{2}-
R_{,\rho}^{2}}(RR_{,T}(\psi_{,T}^{2}+\psi_{,\rho}^{2})-2RR_{,\rho}\psi_{,T}\psi_{,\rho}+R_{,T}
R_{,\rho\rho}-R_{,\rho} R_{,T\rho}),
\end{eqnarray}
where the commas denote differentiation.



\end{document}